\documentstyle[aps,prl,epsfig,floats]{revtex} 
 
\begin{document} 
\draft 
\wideabs{ 
 
\title{Midgap edge states and pairing symmetry of 
       quasi-one-dimensional organic superconductors} 
 
\author{K. Sengupta, Igor \v{Z}uti\'c, Hyok-Jon Kwon,  
       Victor M. Yakovenko, and S. Das Sarma} 
 
\address{Department of Physics and Center for Superconductivity, 
  University of Maryland, College Park, Maryland 20742-4111} 
 
\date{{\bf cond-mat/0010206}, v.1 October 15, 2000, v.2 January 17,
     2001, v.3 February 28, 2001}

\maketitle 
 
\begin{abstract}%
  The singlet $s$-, $d$- and triplet $p$-wave pairing symmetries in
  quasi-one-dimensional organic superconductors can be experimentally
  discriminated by probing the Andreev bound states at the sample
  edges. These states have the energy in the middle of the
  superconducting gap and manifest themselves as a zero-bias peak in
  tunneling conductance into the corresponding edge. Their existence
  is related to the sign change of the pairing potential around the
  Fermi surface.  We present an exact self-consistent solution of the
  edge problem showing the presence of the midgap states for
  $p_x$-wave superconductivity.  The spins of the edge state respond
  paramagnetically to a magnetic field parallel to the vector $\bf d$
  that characterizes triplet pairing.
\end{abstract} 
 
}

\section{Introduction}

Quasi-one-dimensional (Q1D) conductors of the $\rm(TMTSF)_2X$ family
\cite{TMTSF} (the Bechgaard salts) are the first organic materials
where superconductivity was discovered twenty years ago with
$T_c\approx1$ K \cite{Jerome}.  Abrikosov proposed that the
superconductivity is $p$-wave triplet \cite{Abrikosov82}, because it
is suppressed by nonmagnetic impurities \cite{Choi82}.  Gor'kov and
J\'erome observed that the upper critical magnetic field $H_{c2}$
exceeds the Pauli paramagnetic limit, which is also a signature of
triplet superconductivity \cite{Gorkov}.  Recent data show that
$H_{c2}$ exceeds the Pauli limit by a factor greater than 4
\cite{Lee97}.  Another signature of triplet pairing is that the Knight
shift does not change between the normal and superconducting states
\cite{Lee99}.  However, the temperature dependence of the NMR
relaxation rate \cite{Takigawa87} and analogy with the
high-temperature superconductors led to an alternative proposal of the
$d$-wave symmetry \cite{Emery86}.  It was also proposed that a singlet
Q1D superconductivity can overcome the Pauli paramagnetic limit by
forming the spatially nonuniform Larkin-Ovchinnikov-Fulde-Ferrell
state \cite{Dupuis}.  However, the quantitative analysis of the
experimental data \cite{Lee97} by Lebed {\it et al.}
\cite{Lebed99,Lebed00} did not support this proposal and favored
triplet pairing.  The $f$-wave \cite{f-wave} was also proposed
recently.  So the pairing symmetry in the Bechgaard salts remains
hotly debated.

In this paper, we propose a phase-sensitive method to distinguish
experimentally between the $s$-, $p$-, and $d$-wave symmetries.  We
employ a relation between sign change of the superconducting pair
potential around the Fermi surface and existence of the surface
Andreev bound states, discovered for $p$-wave by Buchholtz and
Zwicknagl \cite{Buchholtz81} and for $d$-wave by Hu \cite{Hu}.  For
different superconducting symmetries, we determine which edges of
$\rm(TMTSF)_2X$ must have the Andreev bound states.  The energy of
these states is in the middle of the superconducting gap, thus they
can be observed in tunneling experiments as zero-bias conductance
peaks \cite{Buchholtz81,Tanaka,experiment}.  We also obtain an exact
self-consistent solution of the edge problem for a $p_x$-wave Q1D
superconductor by mapping it onto the kink soliton solution for a 1D
charge-density wave \cite{Brazovskii80}.  We show that the spins of
the edge states should exhibit a strong paramagnetic response to a
magnetic field parallel to the polarization vector $\bf d$ of the
triplet pairing and propose the corresponding experiment.  All
calculations are performed at zero temperature.

\section{Q1D superconductivity}

Classification of superconducting pairing symmetry is particularly
simple for a 1D electron gas.  Its Fermi surface consists of two
points $\pm k_{\rm F}$.  Let us introduce the operators
$\hat{\psi}_\sigma^{\alpha}$ of the right ($\alpha$=R) and left
($\alpha$=L) moving electrons with the momenta close to $\pm k_{\rm
  F}$ and the spin $\sigma$=$\uparrow,\downarrow$.  The Cooper pairing
can be either singlet
$\langle\hat{\psi}_\sigma^{\alpha}\hat{\psi}_{\sigma'}^{\bar\alpha}\rangle
\propto\epsilon_{\sigma\sigma'}\Delta^{\alpha}
=i\hat\sigma^{(y)}_{\sigma\sigma'}\Delta^{\alpha}$ or triplet
$\langle\hat{\psi}_\sigma^{\alpha}\hat{\psi}_{\sigma'}^{\bar\alpha}\rangle
\propto i\hat\sigma^{(y)}({\bf d}\cdot\hat{\bbox{\sigma}})
\Delta^{\alpha}$.  Here $\bar\alpha$=L,R for $\alpha$=R,L;
$\epsilon_{\sigma\sigma'}$ is the antisymmetric metric tensor, and
$\hat{\mbox{\boldmath$\sigma$}}$ are the Pauli matrices acting in the
spin space; {\bf d} is a unit vector of polarization of the triplet
state. Since the fermion operators anticommute, the superconducting
pair potential has either the same ($\Delta^{\rm R}=\Delta^{\rm L}$)
or the opposite ($\Delta^{\rm R}=-\Delta^{\rm L}$) signs at the two
Fermi points for the singlet or triplet pairing.
 
The real $\rm(TMTSF)_2X$ materials are three-dimensional (3D) crystals
consisting of parallel chains.  In the tight-binding approximation,
the electron energy dispersion (measured from the Fermi energy) can be
written as \cite{Yamaji}
\begin{equation} 
\epsilon({\bf k})=v_{\rm F}(|k_x|-k_{\rm F})-2t_b\cos(k_yb)-2t_c\cos(k_zc). 
\label{band} 
\end{equation} 
In the right-hand side of Eq.\ (\ref{band}), the first term represents
the dispersion along the chains, linearized near the Fermi energy with
a Fermi velocity $v_{\rm F}$.  The two other terms describe electron
tunneling between the chains in the $y$ and $z$ directions with the
amplitudes $t_b$ and $t_c$.  ${\bf k}=(k_x,k_y,k_z)$ is the 3D
electron momentum, $b$ and $c$ are the lattice spacings in the $y$ and
$z$ directions, and $\hbar=1$.
 
\begin{figure} 
\centerline{\psfig{file=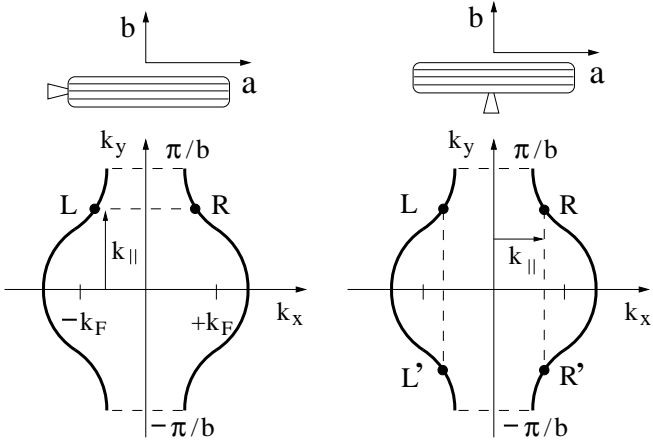,width=\linewidth,angle=0}} 
\caption{ Top: $\rm(TMTSF)_2X$ samples with the lines indicating 1D
  chains.  The left and right panels sketch tunneling along the $a$
  and $b$ axes.  Bottom: The Fermi surface of $\rm(TMTSF)_2X$,
  sketched with a greatly exaggerated warping in the $k_y$ direction.
  Reflection from the edge perpendicular (parallel) to the chains
  changes electron momentum from L to R (R' to R and L' to L), as
  shown in the left (right) panel.}
\label{fig1} 
\end{figure}

The Fermi surface corresponding to Eq.\ (\ref{band}) consists of two
disconnected sheets, sketched in Fig.\ \ref{fig1} with a greatly
exaggerated warping in the $k_y$ direction.  In the simplest case, the
superconducting pair potential $\Delta({\bf k})$ is equal to a
constant $\Delta^{\alpha}$ on a given sheet $\alpha$ of the Fermi
surface, and $\Delta^{\rm R}=\pm\Delta^{\rm L}$ for the singlet or
triplet pairing, respectively.  In both cases, the superconducting gap
has no nodes on the Fermi surface.  These two symmetries can be called
$s$- and $p_x$-waves.  Other symmetries will be discussed at the end
of the paper (see Fig.\ \ref{fig3}).
 
Electron eigenstates of the energies $E_n$ are described in a
superconductor by the Bogolyubov-de Gennes (BdG) wave functions
$\Psi_n=e^{i{\bf r}\cdot{\bf k}_{\rm F}}
[u_{n,\sigma}(x),\epsilon_{\sigma'\bar\sigma'}v^{\bar\sigma'}_n(x)]$,
where $u_{n,\sigma}$ and $v_n^{\bar\sigma'}$ are the electron-like and
hole-like components with the spins $\sigma$ and $\bar\sigma'$, and
$\bar\sigma'=\downarrow,\uparrow$ is the spin index opposite to
$\sigma'=\uparrow,\downarrow$.  The 3D Fermi momenta ${\bf k}_{\rm F}$
belong to the warped Fermi surface shown in Fig.\ \ref{fig1}.  Near
the Fermi surface $\alpha$, the wave functions satisfy the linearized
BdG equation \cite{de89,Andreev64} \FL
\begin{equation} 
  \left(\!\!\begin{array}{cc} 
  -i\alpha v_{\rm F}\partial_x & 
  (\hat{\bbox{\sigma}}\cdot{\bf d})\,\Delta^\alpha(x) \\ 
  (\hat{\bbox{\sigma}}\cdot{\bf d})\,\Delta^{\alpha*}(x) & 
  i\alpha v_{\rm F}\partial_x 
  \end{array}\!\!\right) 
  {u_n^\alpha \choose v_n^\alpha} 
  =E_n{u_n^\alpha \choose v_n^\alpha}, 
\label{BdG} 
\end{equation} 
where $\alpha v_{\rm F}=\pm v_{\rm F}$ for $\alpha$=R,L.  The term
$(\hat{\bbox{\sigma}}\cdot{\bf d})$ is present for triplet
superconductivity and absent for the singlet one.  It operates on the
spin indices of the components $u$ and $v$.

In general, the vector $\bf d$ is a function of the position on the
Fermi surface, e.g.\ in $^3$He-A and $^3$He-B \cite{Leggett}.  The
quantitative analysis \cite{Lebed00} of the experimental data
\cite{Lee97} gives the following information about the components of
the vector $\bf d$ in the $\rm(TMTSF)_2X$ crystal: $d_a\neq0$,
$d_b=0$, and $d_c$ is unknown.  In this paper, we make the simplest
assumption that $\bf d$ is a real vector pointing along the $a$ axis
parallel to the chains (the so-called polar state \cite{Leggett}).  If
we select the spin quantization axis along {\bf d}, then the
4$\times$4 matrix equation (\ref{BdG}) decouples into two 2$\times$2
matrix equations for the wave functions
$(u_\sigma,\epsilon_{\sigma\bar{\sigma}}v^{\bar{\sigma}})$:
\begin{equation}
  \left(\!\!\begin{array}{cc} 
  -i\alpha v_{\rm F}\partial_x & \sigma\Delta^\alpha(x) \\ 
  \sigma\Delta^{\alpha*}(x) & i\alpha v_{\rm F}\partial_x 
  \end{array}\!\!\right) 
  {u_{n,\sigma}^\alpha \choose \sigma v_n^{\alpha,\bar\sigma}}
  =E_n{u_{n,\sigma}^\alpha \choose \sigma v_n^{\alpha,\bar\sigma}}.
\label{E_n} 
\end{equation} 
Here the index $\sigma=\uparrow,\downarrow$ takes the values $\pm$
when used as a coefficient.  It is present in the off-diagonal term
$\sigma\Delta^\alpha$ only for triplet, but not for singlet pairing.
To simplify equations, we omit the spin index $\sigma$ in Secs.\
\ref{Sec:Edge} and \ref{Sec:Tunneling} and restore it in Sec.\
\ref{Sec:Spin}.  Notice that Eqs.\ (\ref{BdG}) and (\ref{E_n}) depend
only on the 1D coordinate $x$, because the 3D dispersion (\ref{band})
in $k_y$ and $k_z$ has been absorbed into the definition of the 3D
Fermi momenta ${\bf k}_{\rm F}$.

\section{Edge states}
\label{Sec:Edge}
 
Let us consider a system occupying the semi-infinite space $x\geq0$
with an impenetrable edge at $x=0$.  When the electron reflects from
the edge specularly, its $k_x$ momentum changes sign, whereas the
other components remain the same.  The electron scatters from the
point ${\bf k}_{\rm F}^{\rm L}$ at the left sheet of the Fermi surface
to the point ${\bf k}_{\rm F}^{\rm R}$ at the right sheet, as shown in
the left panel of Fig.\ \ref{fig1}.  Thus, its BdG wave function
$\Psi$ is a superposition of the R and L terms:
\begin{equation} 
  \Psi=\frac{1}{\sqrt{2}} \left[ 
  e^{i{\bf r}\cdot{\bf k}_{\rm F}^{\rm R}} 
  {u_n^{\rm R}(x)\choose v_n^{\rm R}(x)} 
  -e^{i{\bf r}\cdot{\bf k}_{\rm F}^{\rm L}} 
  {u_n^{\rm L}(x)\choose v_n^{\rm L}(x)} 
  \right]. 
\label{Psi} 
\end{equation} 
We have selected the minus sign in Eq.\ (\ref{Psi}) so that the 
impenetrable boundary condition $\Psi(x=0)=0$ gives 
\begin{equation} 
  u^{\rm R}(0)=u^{\rm L}(0), \qquad v^{\rm R}(0)=v^{\rm L}(0). 
\label{x=0} 
\end{equation} 
 
First let us use a step-function approximation for the pairing
potential: $|\Delta^\alpha(x)|=\Delta_0\theta(x)$.  Then, the plane
waves $[u^\alpha(x),v^\alpha(x)]\propto e^{ik_xx}$ are the
eigenfunctions of Eq.\ (\ref{E_n}) with the energies
$E=\pm\sqrt{(v_{\rm F}k_x)^2+\Delta_0^2}$.  However, the energy is
real also when $k_x$ is imaginary (but not a combination of real and
imaginary parts): $k_x=i\kappa$ and $E=\pm\sqrt{\Delta_0^2-(v_{\rm
F}\kappa)^2}$.  For $\kappa>0$, this solution describes an electron
eigenfunction localized near the edge at $x=0$:
$[u^\alpha(x),v^\alpha(x)]\propto e^{-\kappa x}$.  Because
$u^\alpha/v^\alpha=\Delta^\alpha/(\alpha iv_{\rm F}\kappa+E)$, the
boundary condition (\ref{x=0}) can be satisfied only for $p_x$-wave
with $\Delta^{\rm R}=-\Delta^{\rm L}$, but not for $s$-wave with
$\Delta^{\rm R}=\Delta^{\rm L}$.  Thus, in the $p_x$ case, there is an
edge electron state with the energy in the middle of the
superconducting gap: $E=0$, and the localization length is equal to
the coherence length: $1/\kappa=v_{\rm F}/\Delta_0$.
 
The step-function approximation does not take into account the BdG 
self-consistency condition 
$\Delta^\alpha(x)=g\sum_nu^\alpha_n(x)v^{\alpha*}_{n}(x)$, where $g$ 
is the effective coupling constant \cite{g}, and the sum is taken over 
all occupied states with $E_n<0$ (at zero temperature).  To solve the 
problem, let us extend the wave function (\ref{Psi}) from the positive 
semispace $x>0$ to the full space $-\infty<x<\infty$.  Let us define 
$[u(x),v(x)]=[u^{\rm R}(x),v^{\rm R}(x)]$ and $\Delta(x)=\Delta^{\rm 
  R}(x)$ for $x>0$, and $[u(x),v(x)]=[u^{\rm L}(-x),v^{\rm L}(-x)]$ 
and $\Delta(x)=\Delta^{\rm L}(-x)$ for $x<0$.  Because of the boundary 
condition (\ref{x=0}), the wave function $[u(x),v(x)]$ is continuous 
at $x=0$ and satisfies a single BdG equation for $-\infty<x<\infty$ 
with the BdG self-consistency condition: 
\begin{eqnarray} 
  &&\left(\begin{array}{cc} 
    -iv_{\rm F}\partial_x & \Delta(x) \\ 
    \Delta^*(x) & +iv_{\rm F}\partial_x 
    \end{array}\right) 
    {u_n(x) \choose v_n(x)} 
    =E_n{u_n(x) \choose v_n(x)}, 
\label{E_n'} \\ 
  &&  \Delta(x)=g\sum_nu_{n}(x)v^*_n(x). 
\label{uv*} 
\end{eqnarray} 
 
Eqs.\ (\ref{E_n'}) and (\ref{uv*}) coincide with the exactly solvable
equations describing 1D charge-density wave in polyacetylene
\cite{Brazovskii80}.  The $p_x$-wave problem, where $\Delta(x)$
changes sign: $\Delta(+\infty)=-\Delta(-\infty)$, maps onto the kink
soliton solution \cite{Brazovskii80}:
\begin{eqnarray} 
  && \Delta(x)=i\Delta_0\,{\rm tanh}(\kappa x); 
  \label{scs} \\  
  && E_0=0, \quad {u_0(x) \choose v_0(x)}= 
     \frac{\sqrt{\kappa}}{2\cosh(\kappa x)}{1\choose-1}; 
  \label{E_0} \\  
  && E_k=\pm\sqrt{v_{\rm F}^2k^2+\Delta_0^2},  
  \label{E_k}\\  
  && {u_k(x) \choose v_k(x)}=\frac{e^{ikx}}{2E_k\sqrt{L_x}}  
     {E_k + v_{\rm F}k + \Delta(x) \choose E_k -v_{\rm F}k -\Delta(x)}, 
\label{uv_k} 
\end{eqnarray} 
where $L_x$ is the length of the sample along the chains.  One can
check explicitly that solution (\ref{scs})--(\ref{uv_k}) satisfies
Eqs.\ (\ref{E_n'}) and (\ref{uv*}) \cite{filling}.  It also has the
property of supersymmetry \cite{Goldbart}.  The localized electron
state (\ref{E_0}) with $E_0=0$ corresponds to the Andreev edge state
in the $p_x$-wave superconductor.  In the $s$-wave case, where
$\Delta(x)$ does not change sign: $\Delta(+\infty)=\Delta(-\infty)$,
the solution of Eqs.\ (\ref{E_n'}) and (\ref{uv*}) gives a uniform
$\Delta(x)$, which does not have bound states.  The existence of the
midgap state in the case where $\Delta(x)$ changes sign is guaranteed
by the index theorem and does not depend on the detailed functional
form of the pair potential \cite{index}.

BdG states are described by the operators
$\hat\Psi=u\hat\psi+v^*\hat\psi^\dagger$.  The expectation value of
electric charge $\rho$ in the edge states is zero:
$\rho\propto|u_0|^2-|v_0|^2=0$.  For a 1D $p_x$-wave superconductor
with only one species of spin, Eqs.\ (\ref{Psi}) and (\ref{E_0}) imply
that the two edge states at the opposite ends are described by the
Majorana operators of the opposite parity:
$\hat\Psi^\dagger=\pm\hat\Psi$ \cite{Kitaev}.  There was a proposal to
use such edge Majorana fermions for quantum computing \cite{Kitaev}.
However, in a Q1D $p_x$-wave superconductor, the midgap states with
different momenta ${\bf k}_\|$ parallel to the edge and spins $\sigma$
form a degenerate continuum with $\hat\Psi_{{\bf
k}_\|,\sigma}^\dagger=\pm\hat\Psi_{-{\bf k}_\|,\bar\sigma}$
\cite{CST}.

\section{Tunneling}
\label{Sec:Tunneling}

Let us consider electron tunneling between the superconducting
$\rm(TMTSF)_2X$ and a normal metallic tip.  The tunneling junction can
be modeled as two semi-infinite regions, normal (N) and
superconducting (S), with a flat interface between them.  Following
Refs.\ \cite{Tanaka,Blonder82,Zutic-Valls}, we solve the BdG equations
in the ballistic regime assuming specular reflection and the
translational invariance parallel to the interface \cite{regime}.  To
make the problem analytically tractable, we use the step-function
approximation for the pair potential.  At the interface, we impose the
boundary conditions $\Psi_{\rm N}=\Psi_{\rm S}$ and $\hat{v}_{\rm
N}\Psi_{\rm N}=\hat{v}_{\rm S}\Psi_{\rm S}+2i{\mathcal H}\Psi_{\rm N}$
\cite{Zutic-DasSarma}, where $\hat{v}_{\rm N,S}$ are the components of
the velocity operators perpendicular to the interface in metal and
superconductor, and $\mathcal H$ is the strength of the interface
barrier.  From the solution of the BdG equations, we find the
probabilities $B(E,{\bf k}_\|)$ and $A(E,{\bf k}_\|)$ of the normal
and Andreev \cite{Andreev64} reflections as functions of the electron
energy $E$ and momentum ${\bf k}_\|$ parallel to the interface.  They
determine the dimensionless conductance $G=1+A-B$ in the formula for
the electric current through the contact \cite{Blonder82}:
\begin{eqnarray} 
  I=\frac{2eS}{h}\int\frac{d^2k_\|\,dE}{(2\pi)^2}\, 
  [f(E-eV)-f(E)]\,G(E,{\bf k}_\|). 
\label{IV} 
\end{eqnarray} 
Here $S$ is the contact area, $V$ the bias voltage, $f(E)$ the Fermi
function, $e$ the electron charge, and $h$ the Planck constant.  It
follows from Eq.\ (\ref{IV}) that the differential conductance of the
contact at zero temperature,
\begin{eqnarray} 
  \overline G(V)=\frac{dI}{dV}=\frac{2e^2S}{h} 
  \int\frac{d^2k_\|}{(2\pi)^2}\,G(eV,{\bf k}_\|), 
\label{GV} 
\end{eqnarray} 
is proportional to the average over ${\bf k}_\|$ of the dimensionless
conductance $G$ \cite{cone}.  The latter is determined by the
transmission coefficient $T$ at a given ${\bf k}_\|$
\cite{Tanaka,Zutic-DasSarma}:
\begin{equation} 
  G_\pm = T\,\frac{1 + T\,|\Gamma|^2 + (T-1)\,|\Gamma|^4} 
  {|1 \pm (T-1)\,\Gamma^2\,|^2}, 
\label{gk} 
\end{equation} 
where 
\begin{eqnarray} 
&& \Gamma(E)=\left\{ 
  \begin{array}{ll} 
[E-{\rm sgn}(E)\sqrt{E^2-\Delta_0^2}]/\Delta_0, & |E|\geq\Delta_0, \\ 
(E-i\sqrt{\Delta_0^2-E^2})/\Delta_0, &  |E|\leq\Delta_0, 
  \end{array} \!\! \right. 
\label{Gamma} \\ 
&& T=4 v_{N} v_{S}/[(v_{N}+v_{S})^2+4{\mathcal H}^2]. 
\label{T} 
\end{eqnarray} 
The $\pm$ sign in Eq.\ (\ref{gk}) is the relative sign of the pair
potentials for the two branches of BdG quasiparticles involved in
tunneling.  For tunneling along the chains, the two branches
correspond to the points L and R in the left panel of Fig.\ 
\ref{fig1}, and the sign in Eq.\ (\ref{gk}) is ${\rm sgn}(\Delta^{\rm
  R}\Delta^{\rm L})$: $+$ for $s$-wave and $-$ for $p_x$-wave.
Averaging in Eq.\ (\ref{GV}) is performed taking into account that
$v_N$ and $v_S$ in Eq.\ (\ref{T}) and $\Delta_0$ in Eq.\ (\ref{Gamma})
may depend on ${\bf k}_\|$.

\begin{figure} 
\centerline{\psfig{file=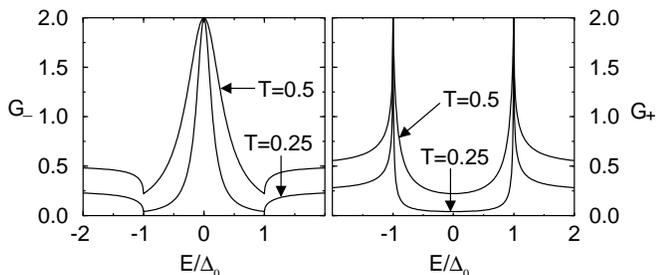,width=\linewidth,angle=0}} 
\caption{ Dimensionless conductances $G_-$ (left panel) and $G_+$
  (right panel) given by Eq.\ (\ref{gk}) are plotted versus energy $E$
  for the transmission coefficients $T=0.5$ and 0.25. $G_+$ and $G_-$
  correspond to the cases where the superconducting pairing potential
  has the same or the opposite signs for the two branches of BdG
  quasiparticles involved in tunneling (the points L and R, R' and R,
  L' and L in Fig.\ \ref{fig1}).}
\label{fig2} 
\end{figure} 
 
As follows from Eq.\ (\ref{gk}), $G_+$ and $G_-$ coincide for a fully
transparent interface ($T=1$): $G_+=G_-=1+|\Gamma|^2$
\cite{Blonder82}.  However, typically $T<1$, both because of the
barrier potential ${\mathcal H}$ and the mismatch of the normal Fermi
velocities $v_N\neq v_S$ in metal and superconductor in Eq.\ (\ref{T})
\cite{Zutic-DasSarma}.  At low interface transparency $T\ll1$, $G_-$
and $G_+$ behave as shown in the left and right panels of Fig.\
\ref{fig2} for $T=0.5$ and $T=0.25$.  Inside the energy gap, where
$|\Gamma|=1$, $G_-(E)$ has a Lorentzian shape with the maximum of 2 at
$E=0$, the width proportional to $T$, and the minimum proportional to
$T^2$ at $|E|=\Delta_0$:
\begin{equation}
  G_-(E)=\frac{T^2/2(1-T)}{(E/\Delta_0)^2+T^2/4(1-T)},
  \quad |E|\leq\Delta_0.
\label{G-}  
\end{equation}
$G_+(E)$ shows the opposite behavior: a minimum proportional to $T^2$
at $E=0$ and the maxima of 2 at $|E|=\Delta_0$:
\[
  G_+(E)=\frac{T^2/2(1-T)}{1-(E/\Delta_0)^2+T^2/4(1-T)},
  \quad |E|\leq\Delta_0.
\]
Both $G_+$ and $G_-$ approach the normal-state conductance $T$ at
$|E|\gg\Delta_0$.
 
\begin{figure}[b]
\centerline{\psfig{file=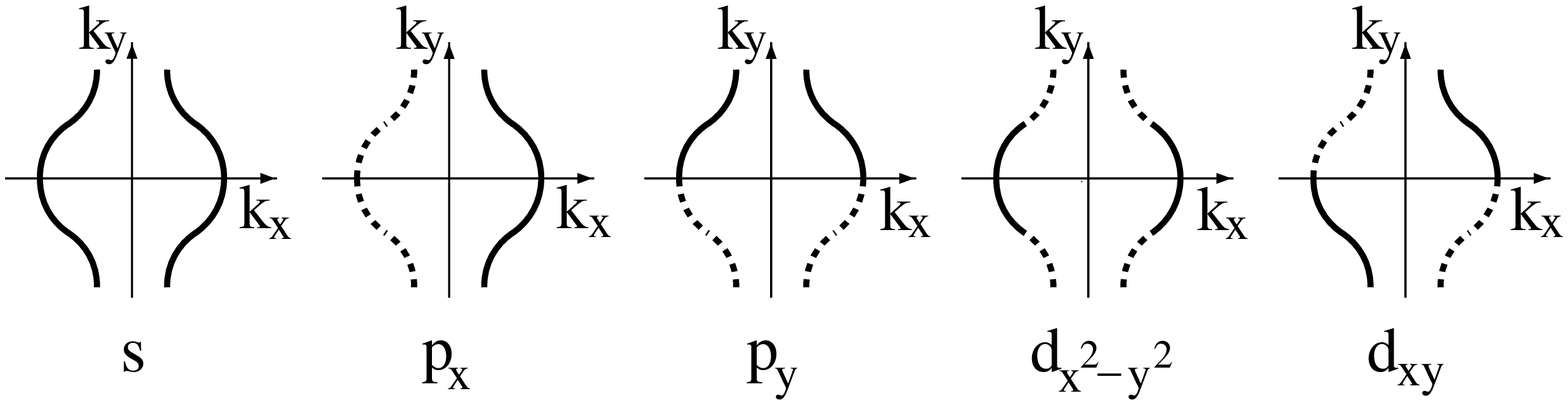,width=\linewidth,angle=0}} 
\caption{ 
  Different symmetries of the pairing potential $\Delta({\bf k})$ in a
  Q1D superconductor. The solid and dotted lines represent the
  portions of the Fermi surface with the opposite signs of the pairing
  potential.}
\label{fig3} 
\end{figure}

The zero-bias conductance peak (ZBCP), shown in the left panel of
Fig.\ \ref{fig2}, is a manifestation of the midgap Andreev bound
states.  They exist at those edges where momentum reflection from the
edge connects the points on the Fermi surface with opposite signs of
the superconducting pair potential.  As shown in the left and right
panels of Fig.\ \ref{fig1}, reflection from the edge perpendicular to
the chains connects L to R, and reflection from the edge parallel to
the chains connects R' to R and L' to L.  By comparing the signs of
the pair potential at these points for the superconducting symmetries
\cite{Lebed00} listed in Table \ref{table} and sketched in Fig.\
\ref{fig3}, we determine whether ZBCP must be present in tunneling
into those edges.  Comparison of Table \ref{table} with the experiment
should uncover the superconducting symmetry of $\rm(TMTSF)_2X$.
 
\begin{table} 
\caption{ 
  Presence (Yes) or absence (No) of a zero-bias conductance peak in
  electron tunneling along the $a$ and $b$ axes (see the top left and
  right panels in Fig.\ \ref{fig1}) for different symmetries of the
  superconducting pairing potential $\Delta({\bf k})$
  \protect\cite{a,s=d}.}
\begin{tabular}{cccc} 
  Symmetry & $\Delta({\bf k})$  & $a$-axis ZBCP & $b$-axis ZBCP \\ \hline 
  $s$ &  const & No & No  \\ 
  $p_x$ & $\sin(k_xa)$ & Yes & No  \\ 
  $p_y$ & $\sin(k_yb)$ & No & Yes  \\ 
  $d_{x^2-y^2}$ & $\cos(k_yb)$ & No & No  \\ 
  $d_{xy}$ & $\sin(k_xa)\sin(k_yb)$ & Yes & Yes  
\end{tabular} 
\label{table} 
\end{table}

\section{Spin response}
\label{Sec:Spin}

Now let us discuss the spin response of the edge states in a triplet
superconductor subject to an external magnetic field $\bf H$.  (Here
we do not consider orbital effects of the magnetic field, such as the
Meissner effect.)  In this case, the matrix in Eq.\ (\ref{BdG}) should
be replaced by the following matrix:
\begin{equation} 
  \left(\!\!\begin{array}{cc} 
  -i\alpha v_{\rm F}\partial_x - \mu_B\,({\bf H\cdot\bbox{\sigma}}) & 
  (\hat{\bbox{\sigma}}\cdot{\bf d})\,\Delta^\alpha(x) \\ 
  (\hat{\bbox{\sigma}}\cdot{\bf d})\,\Delta^{\alpha*}(x) & 
  i\alpha v_{\rm F}\partial_x  - \mu_B\,({\bf H\cdot\bbox{\sigma}})
  \end{array}\!\!\right),
\label{H} 
\end{equation}
where $\mu_B$ is the Bohr magneton, and the electron $g$-factor is 2.

If $\bf H\|d$, then, by selecting the spin quantization axis along
{\bf d}, the 4$\times$4 matrix equation (\ref{H}) is decoupled into
two 2$\times$2 matrix equations similar to Eq.\ (\ref{E_n}) for the
wave functions
$(u_\sigma,\epsilon_{\sigma\bar{\sigma}}v^{\bar{\sigma}})$ with the
matrix
\begin{equation}
  \left(\!\!\begin{array}{cc} 
  -i\alpha v_{\rm F}\partial_x - \sigma\mu_B H
  & \sigma\Delta^\alpha(x) \\ 
  \sigma\Delta^{\alpha*}(x)
  & i\alpha v_{\rm F}\partial_x  - \sigma\mu_B H
  \end{array}\!\!\right).
\label{H||d} 
\end{equation} 
For a given spin projection $\sigma$, the magnetic field $H$ enters
Eq.\ (\ref{H||d}) as a unity matrix and simply shifts the spectrum
(\ref{E_0}) and (\ref{E_k}) by $-\sigma\mu_B H$.  Thus, the energies
of the up and down spin states become split by $\mp\mu_B H$, including
the midgap state: $E_0=\mp\mu_B H$.  Because this state is
half-filled, the edge states with the spin parallel (antiparallel) to
the magnetic field become completely occupied (empty).  This generates
spin $\hbar/2$ and magnetic moment $\mu_B$ at the end of each chain.
Such a giant magnetic moment was predicted by Hu \cite{Hu,Hu-moment}
for the edge states in a singlet $d$-wave superconductor.  In a
triplet superconductor, the effect is similar, but anisotropic.

Indeed, suppose now that $\bf H\perp d$.  In this case, it is
convenient to select the spin-quantization axis $\hat z$ along ${\bf
H}$ and the $\hat x$ axis along ${\bf d}$.  Then Eq.\ (\ref{H})
separates into two 2$\times$2 matrix equations similar to Eq.\
(\ref{E_n}) for the wave functions
$(u_\sigma,\epsilon_{\bar{\sigma}\sigma}v^\sigma)$ with the matrix
\begin{equation}
  \left(\!\!\begin{array}{cc} 
  -i\alpha v_{\rm F}\partial_x - \sigma\mu_B H
  & \Delta^\alpha(x) \\ 
  \Delta^{\alpha*}(x)
  & i\alpha v_{\rm F}\partial_x + \sigma\mu_B H
  \end{array}\!\!\right).
\label{H_|_d} 
\end{equation}
The magnetic field $H$ can be eliminated from Eq.\ (\ref{H_|_d}) by
adjusting the Fermi momenta for the up and down spin states:
$k_{F,\sigma}=k_F+\sigma\mu_B H/v_F$.  Thus, the energy spectrum of
the system remains the same as in Eqs.\ (\ref{E_0}) and (\ref{E_k}).
Particularly, the energy of the midgap state does not split: $E_0=0$,
thus no unbalanced spin and magnetic moment are generated on the edge.

We see that the edge spin response of a triplet superconductor is
opposite to its bulk spin response.  It is well known \cite{Leggett}
that the bulk spin susceptibility for $\bf H\perp d$ is the same as in
the normal state, whereas for $\bf H\| d$ it vanishes at zero
temperature.  For the edge states, the spin response vanishes for $\bf
H\perp d$ and is paramagnetic for $\bf H\| d$.  Nominally, the edge
spin susceptibility is infinite, because, formally, an infinitesimal
magnetic field can completely polarize the edge spins.  We can only
estimate the maximal magnetic moment, which is
$\mu_B=9.3\times10^{-24}$ A~m$^2$ per chain or
$\mu_B/bc=9\times10^{-6}$ $\mu$A per unit area of the edge, where
$b=0.77$ nm and $c=1.35$ nm \cite{Yamaji}.

\begin{figure}[t]
\centerline{\psfig{file=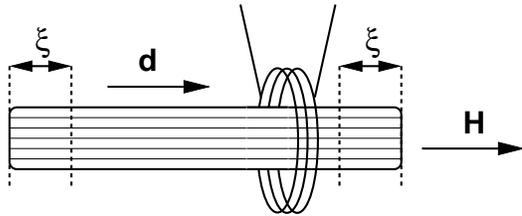,width=0.8\linewidth,angle=0}}
\caption{Schematic experimental setup to measure magnetic
  susceptibility of the edge states localized at the ends of the
  chains.}
\label{fig4}
\end{figure}

The generation of paramagnetic moment by the edge states for $\bf
H\|d$ could be observed experimentally by measuring magnetic
susceptibility with a coil as shown in Fig.\ \ref{fig4}, where we
assume that $\bf d$ is directed along the chains.  In the bulk, far
from the edges, the susceptibility should be diamagnetic, because of
the orbital Meissner effect \cite{Meissner} and vanishing spin bulk
susceptibility for $\bf H\|d$.  However, when the coil is moved toward
the sample end, the susceptibility should change sign and become
paramagnetic because of the edge states.  They are localized within
the coherence length $\xi=\hbar v_F/\Delta_0=0.6$ $\mu$m, where we
used $\Delta_0=0.22$ meV, $v_F=t_aa/\sqrt{2}\hbar=190$ km/s,
$t_a=0.25$ eV, and $a=0.73$ nm \cite{Yamaji}.  The effect should
depend on the coil orientation relative to the vector $\bf d$.
Therefore, this experiment could confirm the existence of the edge
states and the pairing symmetry in the (TMTSF)$_2$X superconductors.

\section{Conclusions}

We have constructed an exact analytical self-consistent solution of
the edge problem for a $p_x$-wave Q1D superconductor by mapping it
onto the kink soliton solution for a 1D charge-density wave.  The edge
electron midgap states exist when the pairing potential has opposite
signs at the different parts of the Fermi surface connected by
momentum reflection from the edge.  These states manifest themselves
as zero-bias peaks in tunneling conductance.  Thus, the pairing
symmetry of the Q1D superconductors can be determined by tunneling
into the edges perpendicular and parallel to the chains.  The spins of
the edge state respond paramagnetically to a magnetic field parallel
to the vector $\bf d$ that characterizes triplet pairing, generating
the magnetic moment $\mu_B$ per chain.

The $\rm(TMTSF)_2X$ materials are expected to have electron edge
states also in the magnetic-field-induced spin-density-wave phase
(FISDW), which exhibits the quantum Hall effect \cite{FISDW}.  Those
states are chiral and have dispersion inside the energy gap.  The
midgap states discussed in the present paper also acquire chiral
dispersion due to the orbital effect of a magnetic field, similar to
cuprates \cite{Sauls97}.  Another interesting example of the edge
states in a 1D electron gas was studied theoretically in Ref.\
\cite{Gogolin} for the Luther-Emery fermions in the bosonized
representation.
 
K.S., H.J.K., and V.M.Y.\ were supported by the Packard Foundation and
NSF Grant No.\ DMR-9815094; I.\v{Z}. and S.D.S. by the US-ONR and
DARPA.

{\it Note added in proofs.}  A more subtle consideration of the
results of Sec.\ V shows that the spin of an edge state is actually
fractional and is equal to $\hbar/4$ (per each end of each chain), not
$\hbar/2$.  Correspondingly, the magnetic moment is $\mu_B/2$, and the
numerical estimates given in Sec.\ V should be multiplied by an
additional factor 1/2.  Derivation of these results will be given in a
separate paper.  The authors are grateful to A.\ Yu.\ Kitaev and D.\
A.\ Ivanov for very illuminating discussions of this subject (see D.\
A.\ Ivanov, cond-mat/9911147).



\begin{references}
\bibitem{g} In the $g$-ology model \cite{Yamaji}, the coupling 
  constants $g$ for singlet and triplet superconductivity are 
  expressed in terms of forward ($g_2$) and backward ($g_1$) 
  scattering amplitudes: $g=g_2\pm g_1$.
\bibitem{filling} The states (\ref{uv_k}) with $E_k<0$ are completely
  filled, whereas the state (\ref{E_0}) with $E_0=0$ is half-filled
  \cite{Hu,Brazovskii80}.
\bibitem{regime} The $\rm(TMTSF)_2X$ materials have high purity and
  the mean free path about 10 $\mu$.
\bibitem{a} $a=\pi/2k_F$ is the crystal lattice period along the
  chains.
\bibitem{s=d} $d_{x^2-y^2}$-wave belongs to the same crystal symmetry
  class as the $s$-wave, so it may be thought of as the extended
  $s$-wave.  That is why both of them do not show ZBCP.  Similarly,
  $f$-wave proposed in Ref.\ \cite{f-wave} belongs to the same class
  as $p_x$-wave and has the same ZBCP.
\end{references}
\end{document}